\documentstyle[preprint,aps]{revtex}

\begin{document}
\draft
\title{
Concerning order and disorder in the ensemble of Cu-O\\
chain fragments in oxygen deficient planes of Y-Ba-Cu-O.
}
\author{
 Gennadi Uimin}
\address{
Institut f\"ur Theoretische Physik der Universit\"at zu K\"oln,\\
Z\"ulpicher Strasse 77, D-50937 K\"oln 41, Germany\\
 and\\
Landau Institute for Theoretical Physics,
Chernogolovka, Moscow District 142432, Russia
}
\date{}

\maketitle

\begin{abstract}
In connection with numerous X-ray and neutron investigations of some high
temperature superconductors (YBa$_2$Cu$_3$O$_{6+x}$ and related compounds)
a non-trivial part of the structure factor, coming from partly disordered
{}~Cu-O-$\dots$-O-Cu~ chain fragments, situated within basal planes,
CuO$_x$, can be a subject of theoretical interest.
Closely connected to such a diffusive part of the structure factor are
the correlation lengths, which are also available in neutron and X-ray
diffraction studies and depend on a degree of oxygen disorder in a basal
plane. The quantitative measure of such a disorder can be associated with
temperature of a sample anneal, $T_q$, at which oxygen in a basal plane
remains frozen-in high temperature equilibrium after a fast quench of a
sample to room or lower temperature. The structure factor evolution with
$\,x\,$ is vizualized in figures after the numerical calculations.
The theoretical approach employed in the paper has been developed for the
orthorhombic state of YBCO.
\end{abstract}
\pacs{74.70.Vy, 64.60.Cn, 64.80.Eb}

\section{Introduction.}
While the regular CuO$_2$ planes are commonly
believed to be responsible for both fascinating phenomena, the
high-$T_c$ superconductivity (HTS) and high-$T_N$ antiferromagnetism,
the oxygen deficient planes of the YBa$_2$Cu$_3$O$_{6+x}$ compounds play
an important role too, taking a control over the mentioned phenomena.
HTS is associated with the orthorhombic state of YBCO which is
consistent with existence of linear chain fragments, ~Cu-O-$\dots$-O-Cu,
in oxygen deficient planes, CuO$_x$.  Even in the case of twinned
crystals the substance consists of macroscopical domains where chain
fragments (CFs) are arranged in chains (conventional notation is O(1)
and O(5) for oxygen positions in and across the chains, respectively).
There are three orthorhombic phases, Ortho-I, II and III, which have
been reliably identified experimentally.  Concerning the Ortho-II and
Ortho-III phases, their identification in neutron and X-ray diffraction
studies \cite{1,2,3,4} has been reported.  The part of the structural
phase diagram in the $x$-$T$ plane is shown qualitatively in Fig.1 (cf
Refs \cite{5,6,7}).  The $T$ coordinate plays the role of temperature of
oxygen equilibrium in the basal plane.  It could be of the order of room
temperature, $T_R$, if a cooling procedure were done slowly.  Otherwise,
it would be the so-called quenching temperature, $T_q$, if a sample were
annealed for a long time and then quenched rapidly from ~$T_q$ to room
or lower temperature.  The role of $T_R$ is important: Oxygen atoms in
deficient planes do not diffuse below ~$T_R$ (cf, however, the elegant
investigation by Jorgensen {\it et al} ~\cite{8}). $T_q$'s have varied
up to 350$^{\circ}$C in experiments by Veal and co-workers (\cite{9,10}
and references therein).  In order to prepare the Ortho-III phase the
samples have been quenched at temperatures above 500$^{\circ}$C
{}~\cite{3}.

As mentioned above the constituents which form the oxygen deficient
plane are the CFs. In the tetragonal phase they are rather short and
could be assumed to be situated at random were it not for evidence of a
superstructure at some low oxygen content, $x\approx 0.35$, as reported
in \cite{11,12} (cf \cite{13}, however).  Most likely, the origin of the
herring-bone arrangement of the very short Cu-O-Cu chain pieces is
simply owing to the Coulomb interaction of such fragments which are
simultaneously quadrupoles (see \cite{14}).

In the Ortho-I phase the distribution of CFs over their lengths is
similar in all the chains.  The single-cell structure is typical for the
right part of the phase diagram (0.75$<\!x\!<$1 at $\,T_q\!\sim\!
T_R$).  Between Tetra and Ortho-I phases in Fig.1 there is the area of
the Ortho-II phase, which simply means a double-cell structure with
oxygen-poor chains alternating oxygen-rich ones.  However, the phase
transition lines, restricting this area, are not strictly determined up
to now: There is still a question about the type of the
Tetra-to-Ortho-II phase transition.  Either the system could undergo the
phase separation, ie, the first order phase transition with phase
co-existence in the form of the "gas"(Tetra)--"liquid"(Ortho-II) mixture
or the phase transition through a growth of Ortho-II domains.  More
acknowledged is the first scenario, see, for instance, \cite{5,6,15,16},
but this conclusion has been done in a frame of a lattice-gas model, the
so-called ASYNNNI model, which ignores intrinsic, ie, fermionic degrees
of freedom of CFs. As for the second scenario the domains of four
various kinds would be locally arranged in the double-cell structures,
oriented along $b$-/$a$-axis and shifted across the chains by a single
period.  Another boundary, confining the Ortho-II phase, would divide
the regions of the Ortho-I and Ortho-II phases were it not for the
possibility for the Ortho-III phase to exist between them.  The
periodicity of the latter corresponds to one oxygen-poor chain following
two oxygen-rich chains.  However, in order to search the Ortho-III phase
experimentally the way of a sample preparation should be indirect.
Actually, a slow cooling is accompanied with a formation of the Ortho-II
phase, which locks the periodicity, making the kinetics of the phase
transformation into the Ortho-III phase to be embarrassing.  To avoid
such a difficulty a slow cooling in the atmosphere of higher oxygen
pressure at the first stage can be switched over to a fast quench from
rather high temperature ($>$500$^{\circ}$C) at the second stage
\cite{3}.  There was a theoretical conjecture concerning a possibility
for long wave-length modulated structures to exist in oxygen deficient
planes (see \cite{18} and references therein).  However, the afore-said
impossibility of oxygen migration at low enough temperatures
($T$$<$$T_R$) makes such hypothetical modulated structures the subject
of an abstract interest.

Note that the Ortho-I phase can be interpreted as a disordered Ortho-II
phase with a short-range order. Certainly, the order parameter of the
Ortho-I phase is of the higher symmetry than that of the Ortho-II phase.

To reveal the regions of stability of Tetra, Ortho-I and Ortho-II phases
and to construct a structural phase diagram theoretically a
two-dimensional lattice-gas model, the ASYNNNI model, has been used as a
prerequisite by de Fontaine and colleagues.  The model presented in
Fig.2 (on the parameters, entering the model, see, for instance,
\cite{5,6,19}) invokes the nearest neighbor oxygen-oxygen repulsion
($V_1$, oxygen in the "corner" configuration), the next-nearest neighbor
attraction ($-V_2$, in-line configuration) and also the next-nearest
repulsion without any Cu ion in between the corresponding oxygen atoms
($V_3$, "vis-\`a-vis" configuration).  As mentioned above the
lattice-gas model ignores a fermionic origin of microscopical properties
of CFs. Also this model is not the best candidate for a proper
description of a structure formation: The superstructure at small
$\,x\,$ is beyond the ASYNNNI model predictions.  Nevertheless, the
lattice-gas approach is reasonable for the orthorhombic side of the
phase diagram and it can be phenomenologically generalized to
incorporate the fermionic degrees of freedom into the effective
interaction of oxygen atoms within CFs. Generalization is based on the
result obtained in the soft X-ray absorption spectroscopy investigation
of untwinned YBa$_2$Cu$_3$O$_7$ and reported in \cite{20}: The
concentration of oxygen holes in the Cu-O chains has been estimated
around 30\%.  This experimental estimate is essential for applying the
charge transfer mechanism discussed theoretically in \cite{21} and
partly based on the experimental investigations \cite{-21,-22} of the
monovalent $Cu$ amount in YBa$_2$Cu$_3$O$_{6+x}$.  A short resum\'e of
the mechanism proposed in \cite{21} is as follows: Existence of the CF,
involving $\,n\!+\!1\,$ Cu$^{2+}$ and $\,n\,$ O atoms, $\,m\,$ of which
are O$^{2-}$ and $\,n\!-\!m\,$ atoms are O$^-$, is consistent with
$\,m\!-\!1\,$ holes transferred from the fragment.  In the limit of very
long CFs ($n\!\gg \!1$), ie, the case investigated in Ref \cite{20}, the
free energy per one oxygen site of the fragment, $f(n,m)$, depends
practically of $\,m/n\,$ and has a minimum around $\,\xi\approx $ 0.70.
It allows to represent the free energy in the following form:
\begin{eqnarray}
\label{1}
f(n,m)=f_n+\frac {a_n}2\left(\frac mn-\xi_n\right)^2
\end{eqnarray}
where $\,f_n$, $a_n$ and $\xi_n$ go to the fixed values when
$n\rightarrow \infty$.  One employs Eq.(\ref{1}) to get a reduced
information about CFs, expressed in terms of their lengths $\,n$:
\begin{eqnarray}
\label{2}
e^{-n\phi(n)/T}=\sum_{m=0}^ne^{-nf(n,m)/T}
\end{eqnarray}
This procedure is convenient to build the statistical mechanics of the
CF ensemble because the configurational entropy is associated with the
lengths of CFs. The reduced free energy, $\phi(n)$, reads at
$\,n\!\gg$1:
\begin{eqnarray}
\label{3}
\phi(n)=f_n-\frac T{2n}\ln\frac{2\pi nT}{a_n}\approx\phi_1+
\frac {\phi_2}n -\frac T{2n}\ln n
\end{eqnarray}
Expression (\ref{3}) could be compared with the analogous one easily
obtained from the lattice-gas model mentioned above.  Actually, the
attraction energy $\,-V_2(n-1)\,$ plays the same role as $\,n\phi(n)\,$
in formation of the CF of length $\,n$.  So, the analogue of
Eq.(\ref{3}) is
\begin{eqnarray}
\label{4}
\tilde\phi(n)=-V_2+\frac {\tilde V_2}n
\end{eqnarray}
Deriving Eq.(\ref{4}) we also include the energy loss, $V_{3-{\rm Cu}}$,
due to the 3-fold coordinated Cu ions at the ends of a CF, which yields
the evident change $\,V_2/n\rightarrow\tilde V_2/n= (V_2+2V_{3-{\rm
Cu}})/n$.  The first terms in the r.h.s. of Eqs.(\ref{3}) and (\ref{4})
are not relevant if oxygen atoms predominantly occupy O(1) positions:
For the orthorhombic state of YBCO, prepared at moderate (not too high!)
$T_q$, O(5) positions were proved \cite{22} to be occupied poorly.  This
is the case for which one assumes the total amount of oxygen to be
fixed, hence, the first, $n$-independent, term of
Eqs.(\ref{3})-(\ref{4}) does not distinguish between various
configurations of CFs. The second terms in Eqs.(\ref{3})-(\ref{4}) are
important and their meaning is clear: CFs would show a tendency to
increase their lengths were it not for the third term in Eq.(\ref{3})
and, in addition, for the configurational entropy of the CF ensemble.
Aforementioned temperature, $T=T_q$, is above, at least, than $\,T_R$.

In order to complete the scheme presented by Eq.(\ref{3}) or (\ref{4})
the following parts of the free energy must be involved into a
theoretical consideration: {\it First}, the configurational energy term
due to various configurations of CFs, and {\it second}, caused by the
interaction of oxygen atoms in different chains, ie, the $V_3$-term.
Partly, this line of attacking the problem has been realized in
\cite{23,24} and subjected to the experimental check by Boucherle {\it
et al} ~\cite{25}.  Such a general theoretical approach as well as some
important applications to the experimentally observable values, reported
already and foreseen, will be presented in the following sections.

There is also a more straightforward approach to the problem of CFs in
the orthorhombic phase of YBCO.  Instead of employing phenomenological
schemes of Eq.(\ref{3}) or (\ref{4}) it would be possible to start with
some microscopical model for an alternating Cu-O CF, which should belong
to a family of strongly correlated fermionic systems despite of its
finite length.  A good candidate could be the Emery model \cite{26}, or
its Kondo-like version \cite{27}, or the 1d model of the Zhang-Rice type
\cite{-27}, derived from the original Emery model: All are well adapted
to a proper description of the alternating CFs. However, to adjust such
models to the reality one needs to invoke some additional short-range
interactions, ie, the nearest Cu-O and O-O interactions, the bare energy
of a charge transfer from chains into CuO$_2$ planes and the energy loss
because of the 3-fold coordinated Cu atoms at both ends of CFs. The
problem seems to be hopeless because of too many energy parameters
invoked in, however, the choice of the model parameters is based only on
few relevant properties of CFs, the most important among them is a true
position of the free energy minimum around $m/n\approx$ 0.7.  The
attempt to apply the Kondo-reduced version of the Emery model has been
done in Ref \cite{21}: In addition to the above-mentioned criterium the
plateau-like behavior of $\,n_{\text h}$ vs $\,x$, imitating the 60 K
$\,T_c$ vs $\,x\,$ plateau in YBCO, has been employed as a crucial point
($n_{\text h}$ the concentration of holes leaving the chains).

Certainly, a big advantage of the microscopical model approach is in a
possibility of direct calculations of such physical quantities as
$n_{\text h}$, $n_{\text{pm}}$, the concentration of paramagnetically
active CFs, studied by polarised neutrons \cite{25}, magnetic
correlations within CFs and so on.  However, the chain, consisting of
more than 9 copper and 8 oxygen atoms, are not available to be treated
by means of the exact numerical diagonalization algorithm by Lanczos
(see the review paper \cite{28} for the technical details of the
method).  Nevertheless, the quantum Monte Carlo algorithm seems to be
promissing in studying of microscopical properties of CFs at non-zero
temperature.

In this paper we mainly follow the line of the generalized lattice-gas
model, whose peculiarities are expressed by Eq.(\ref{3}).

The principal results of the paper are collected in Section III, while
all the important spade-work is done in Section II and Appendixes.
Section II starts with application of the linear programming method for
selecting CFs of optimal lengths.  But the method works reliably only at
rather low temperatures when the contribution of the configurational
entropy can be neglected.  Then a more realistic situation is
considered: The average lengths of chain and vacancy fragments are
calculated as a function of oxygen filling in a chain vs temperature;
the in-chain structure factor as well as the in-chain oxygen
density-density correlation functions are analysed.  In Appendix B we
represent the equations of statistical mechanics of the ensemble of CFs
arranged orthorhombically by means of the distribution of CFs over their
lengths.  In Section III the interchain interaction is incorporated by
means of the high-temperature expansion of the free energy.  The final
results are partly expressed in the analytical form (Section III),
partly visualized after numerical calculations.  They are presented in
Section IV, which also includes the set of plots for some observable
physical quantities, playing the role of the reference set.

Before starting the main part of the paper, let us introduce some
definitions: $N_m$, the total number of chain fragments, consisting of
$\,m+$1 ~Cu atoms and $\,m\,$ O atoms.  It will be convenient,
especially for further application to the statistical mechanics of the
CF ensemble, to incorporate one of the vacant of oxygen Cu-Cu links,
surrounding CF, to be certain, to the right, into it.  These numbers and
corresponding "probabilities" $x_m\!=\!N_m/{\cal N}_{\ell}$ satisfy the
evident constraints:
\begin{eqnarray}
\label{5}
\frac 1{{\cal N}_{\ell}}\sum_{m=0}(m+1)N_m=\sum_{m=0}(m+1)x_m=1
\end{eqnarray}
and
\begin{eqnarray}
\label{6}
S=\frac 1{{\cal N}_{\ell}}\sum_{m=0}N_m=\sum_{m=0}x_m=1-x
\end{eqnarray}
where ${\cal N}_{\ell}$ is the total number of Cu atoms in a chain. $S$
is, evidently, the concentration of oxygen vacancies in a chain.  Many
experiments in YBCO have been performed at the fixed oxygen content:
Because of a possibility for oxygen to fill the chains inhomogeneously
one redefines Eq.(\ref{6}) by taking a sum over the chains:
\begin{eqnarray}
\label{7}
S=\frac 1{{\cal N}_{{\rm ch}}}\sum_{\alpha}\sum_{m=0}x_m^{(\alpha)}=1-x
\end{eqnarray}
where $\alpha$ labels the chains and ${\cal N}_{{\rm ch}}$ is their
total number.
\section{In-chain distribution of CF\lowercase{s} over the lengths.}
\subsection{A hypothetical low-temperature case.}
Using the intrinsic free energy of a CF, Eq.(\ref{3}), let us elucidate
an auxilary problem, namely, what kind of CFs would be preferable if the
configurational entropy as well as the interchain interaction were
neglected.  The free energy of the ensemble of non-interacting CFs (per
one copper site) can be expressed as a {\it linear} function of the set
$\{x_m\}$, which is also subjected to the {\it linear} constraints
(\ref{5})-(\ref{6}):
\begin{eqnarray}
\label{8}
f=\sum_{m=1}m\phi(m)x_m=\sum_{m=1}(u-\tau\ln m)x_m,
\end{eqnarray}
where $u=\phi_2,\;\;\tau=T/2$. $f$ can be minimized by making use the
linear programming method \cite{29}.  In the physics of the CF ensemble
this procedure would reflect the properties of non-interacting
constituents, ie, CFs, if they were supposed to be annealed at rather
low temperature.  Note, however, that such a case has no direct bearing
on the YBCO behavior, because oxygen of deficient planes would be
supposed to be frozen below $\,T_R$.

The linear programming theory says \cite{29} that a function, $f\,$ in
this case, will attain its minimum on a unique vertex of the convex
polytope $\,{\cal P}\,$ defined by constraints (\ref{5})-(\ref{6}).  Any
vertex $\,[i,j]\,$ defined by Eqs.(\ref{5})-(\ref{6}) can be represented
by a pair of non-zero probabilities ($i<j$):
\begin{eqnarray}
\label{9}
x_i=\frac{(j+1)(1-x)-1}{j-i},\;\;x_i=\frac{1-(i+1)(1-x)}{j-i},
\;\;\;x_m=0\;\;\text {if}\;\;m\neq i,\;m\neq j
\end{eqnarray}
Certainly, not all the pairs are the candidates to be vertices, a true
vertex of this particular problem is composed of two {\it non-negative}
probabilities.  This results in the following available set:
\begin{eqnarray}
\label{10}
j>\kappa>i,\;\;\;\kappa=\frac 1{1-x}-1=\frac {1-S}S
\end{eqnarray}
True vertex $\,[i,j]\,$ defines $\,f\{i,j\}$, the function to be
optimized, as follows:
\begin{eqnarray}
\label{11}
f\{i,j\}=(1-x)\left(u-\tau\left(\frac{\kappa-i}{j-i}\ln j+
\frac{j-\kappa}{j-i} \ln i\right)\right)
\end{eqnarray}
It is proved in Appendix A that $\,f\{i,j\}\,$ attains the global
minimum on the manifold of vertices, defined by Eq.(\ref{9}) with
$\,i\,$ and $\,j\,$ both non-zero, on unique vertex
$\,[\kappa_-,\kappa_+]\,$ ($\,\kappa_{\pm}$ are the two nearest integers
to $\kappa$):
\begin{eqnarray}
\label{12}
f\{\kappa_-,\kappa_+\}=(1-x)\Big(u-\tau\big((\kappa-\kappa_-)\ln
\kappa_+ +(\kappa_+-\kappa)\ln \kappa_-\big)\Big)
\end{eqnarray}

However, Eq.(\ref{12}) ignores the $\,[0,m]\,$ vertices: Although
Eq.(\ref{9}) is valid in this case also,
$$
x_0=1-\frac{m+1}mx,\;\;x_m=\frac xm;\;\;\;m>\kappa,
$$
the energy expression of Eq.(\ref{8}) differs from the Eq.(\ref{11})
form:
\begin{eqnarray}
\label{13}
f\{0,m\}=\frac{\kappa}m(1-x)(u-\tau\ln m);\;\;\;m>\kappa
\end{eqnarray}
Formally, $\,f\{0,m\}$ as a function of a real variable ($m\!\rightarrow
\! \mu$) would achieve its minimum at $\mu=\exp (u/\tau+1)$, hence, one
of the nearest integers to $\mu$, $\,\mu_-\,$ and $\,\mu_+$, is the true
candidate to be the minimum point.  To analyse what kind of oxygen
arrangement in CFs is favorable we should compare two competing minima
of Eqs.(\ref{12})-(\ref{13}).  Such an analysis is put into Appendix A
and it says that the global minimum of the free energy $\,f\,$ is
situated on the $\,[0,\mu_{\pm}]\,$ vertex if $\kappa_+\!<\!\mu$ and on
the $\,[\kappa_-,\kappa_+]\,$ vertex if $\kappa_-\!>\!\mu$.  Switching
over the vertices occurs at some $\mu$ between $\kappa_-$ and
$\kappa_+$.

Less rich is the information about "vacancy fragments" (VFs), ie, those
pieces of the chains which are free of oxygen atoms.  Nevertheless, one
can define the average length of such VFs and express it in terms of the
average length of CFs, $\overline {\ell}$, as
\begin{eqnarray}
\label{14}
\overline {\ell_v}=\frac{\displaystyle \sum_{k=0}N_k}{\displaystyle
\sum_{k=1}N_k}=\frac S{S-x_0}=\frac{\overline {\ell}}{\kappa}
\approx\frac{\mu}{\kappa}
\end{eqnarray}
Note, that there is a broad distribution of VFs over their lengths
around $\,\mu/\kappa$.

The conclusions of this subsection are applicable to the ensemble of
non-interacting chain (and vacancy) fragments:
\begin{itemize}
\item
If the oxygen content, $x$, is not close to 1, ie, in the
moderate-$\kappa$ case, the preferable length of CFs is determined by
$\,\mu$.  The average length of VFs could be estimated as
$\,m_v\!\approx\!\mu/\kappa$.
\item
If $\,\kappa_+$ exceeds $\,\mu$, the average length of CFs is determined
by $\,\kappa$.  They are separated by single oxygen vacancies.
\end{itemize}
In the following sections the role of the configurational entropy will
be revealed and the first conclusion of this subsection will be
significantly revised.
\subsection{The role of configurational entropy.}
By the next step we incorporate the configurational entropy term into
the free energy of a 1d chain.  This will be done in order to estimate
the average length of chain fragments, which is available in the neutron
and X-ray diffraction studies (see \cite{1,2,3,4}).  A theoretical
prerequisite is collected in Appendix B where the "probability" set,
$\{x_m\}$, is derived (recall, that the formulas below stand upon
Eq.(\ref{3}) (see also \cite {23,24})):
\begin{eqnarray}
\label{15}
x_m=\frac {{\sqrt m}z^m}{\displaystyle
e^{\Phi}+\sum_{m=1}{\sqrt m}(m+1)z^m} \;\;\;(m\geq 1)
\end{eqnarray}
\begin{eqnarray}
\label{16}
x_0=\frac {e^{\Phi}}{\displaystyle e^{\Phi}+\sum_{m=1}{\sqrt m}(m+1)z^m}
\end{eqnarray}
Certainly, constraint (\ref{5}) is satisfied by
Eqs.(\ref{15})-(\ref{16}) where $\,\Phi\!=\!\phi_2/T$. $\,z\,$ plays the
role of fugacity and could be determined from Eq.(\ref{6}), rewritten in
the following form:
\begin{eqnarray}
\label{17}
S=\frac {\displaystyle e^{\Phi}+\sum_{m=1}{\sqrt m}z^m}
{\displaystyle e^{\Phi}+
\sum_{m=1}{\sqrt m}(m+1)z^m}
\end{eqnarray}
We perform the summation over $\,m\,$ by using substitution
$z\!=\!\exp -1/2m_0$, assuming that $m_0\!\gg \!1$. Then, for example,
in the main order one obtains
\begin{eqnarray}
\label{18}
\sum_{m=1}\sqrt {m}m^nz^m
\rightarrow\sqrt {m_0}m_0^{n+1}
\int_0^{\infty}dx \sqrt{x}x^ne^{-x/2}
=\sqrt {2\pi m_0}m_0^{n+1}(2n+1)!!
\end{eqnarray}
One may check by using the Poisson summation formula that the main order
correction to Eq.(\ref{18}) measured in the Eq.(\ref{18}) units is
expressed through Riemann's zeta function,
$$
\frac 2{(4\pi m_0)^{n+3/2}}\cos\left(\frac n2\pi+\frac 34\pi\right)
\zeta\left(n+\frac 32\right)
$$
and small at any $m_0$ of the real interest.

Following such a summation recipe we get from Eq.(\ref{17}):
\begin{eqnarray}
\label{19}
S=\frac{e^{\Phi}+\sqrt{2\pi}m^{3/2}_0}{e^{\Phi}+3\sqrt{2\pi}m^{5/2}_0
+\sqrt{2\pi}m^{3/2}_0}
\end{eqnarray}
Eq.(\ref{19}) can be properly solved by substitution:
\begin{eqnarray}
\label{20}
m_0^{5/2}=\frac{1-S}S\frac{e^{\Phi}}{3\sqrt{2\pi}}
\end{eqnarray}
which assumes $\,m_0^{3/2}$ to be neglected as compared to $\,e^{\Phi}$
in the r.h.s. of Eq.(\ref{19}).  On the contrary, when $\,S\!\ll \!1$,
$\,e^{\Phi}\!\ll \!m_0^{3/2}$ and the analogue of Eq.(\ref{20}) reads
\begin{eqnarray}
\label{21}
m_0=\frac 1{3S}
\end{eqnarray}

For the average length of CFs, $\overline {\ell}$, one assumes the
following definition:
\begin{eqnarray}
\label{22}
\overline {\ell}=\frac{\displaystyle \sum_{k=1}k\cdot N_k}
{\displaystyle \sum_{k=1}N_k}=\frac{1-S}{S-x_0}
\end{eqnarray}
One may easily check that
\begin{eqnarray}
\label{23}
\overline {\ell}=3m_0,\;\;\;m_0\gg 1
\end{eqnarray}
independently of the $S$-range.  The same estimate can be found after
minimization of $\,mx_m\,$ over $\,m$.  However, such a distribution is
very broad, its half-width exceeds 2$\overline {\ell}$.  This
circumstance makes the situation significantly distinct of that
discussed in subsection II.A.

The temperature dependence of $\,\overline {\ell}\,$ is exponential when
$\,S\,$ is not very close to 1:
\begin{eqnarray}
\label{24}
\overline {\ell}=\left(\frac{1-S}{3\sqrt{2\pi}S}\right)^{2/5}e^{2\Phi/5}
\end{eqnarray}
and practically $T$-independent
\begin{eqnarray}
\label{25}
\overline {\ell}=1/S
\end{eqnarray}
for $\,S\,$ in a close vicinity of 1.

By using Eqs.(\ref{14}),(\ref{16}) and (\ref{19}) the average length of
VFs can be estimated as
\begin{eqnarray}
\label{26}
\overline {\ell_v}=\frac S{S-x_0}=1+\frac{e^{\Phi}}{\sqrt{2\pi}m_0^{3/2}}
\approx 1+3m_0\frac S{1-S}
\end{eqnarray}
At $\,S\!\ll\!1$ $\,\overline {\ell_v}\!\approx\!1$, otherwise
$\,\overline {\ell_v}\!\approx\!\overline {\ell}S/(1-S)$.  This
conclusion, concerning VFs, is similar to that of subsection II.A.
\subsection{In-chain structure factor and density-density correlation
functions.}
Here we derive the equations and suggest the computational recipes of
the in-chain correlations.  Apart from the other values they allow to
compute the in-chain irreducible averages which enter the
high-temperature expansion of the free energy.

We start with a calculation of the Fourier components of the
oxygen-oxygen correlation function, expressed through the distribution
of CFs over their lengths in the following form:
\begin{eqnarray}
\label{27}
B(q)=S\left\{\left(\left(1-\frac 1S
\sum_{m=0}x_m e^{iq(m+1)}\right)^{-1}+\text{c.c.}\right)-1\right\}
\end{eqnarray}
In order to prove Eq.(\ref{27}) one reminds a convention of the chain
fragment definition: If $\{m_1,m_2,m_3,\dots\}$ is the sequence of
lengths of the CFs, then the coordinates of oxygen vacant sites are
$\{m_1+1,m_1+m_2+2,m_1+m_2+m_3+3,\dots\}$.  The Fourier component of the
oxygen density (not averaged!) is simply
\begin{eqnarray}
\label{28}
g(q)=\sum_{n=1}e^{iqn}-\left\{e^{iq(m_1+1)}+e^{iq(m_1+m_2+2)}+\dots\right\}
\end{eqnarray}
The first sum in the r.h.s. of Eq.(\ref{28}) is zero for any non-zero
$\,q\,$ and we omit it because the subject of interest is the
irreducible averages.  The probability to find a CF of the length
$\,m\,$ in the chain fragment ensemble is $\,x_m/\sum_{k=0}x_k=x_m/S$.
Averaging $\,g(q)\,$ over the distribution of the CF lengths according
to the auxilary formula
$$
\langle e^{iqm}\rangle=\frac 1S\sum_{m=0}e^{iqm}x_m
$$
one obtains:
\begin{eqnarray}
\label{29}
\langle g(q)\rangle =-\left<\left\{e^{iq(m_1+1)}+e^{iq(m_1+m_2+2)}+
\dots\right\}\right>=
-\frac {\displaystyle\frac 1S\sum_{m=0}e^{iq(m+1)}x_m}
{\displaystyle 1-\frac 1S\sum_{m=0}e^{iq(m+1)}x_m}
\end{eqnarray}
According to a definition the in-chain structure factor reads
\begin{eqnarray}
\label{30}
B(q)=\frac 1{{\cal N}_{\ell}}\left<\sum_{r_1,r_2}n_{r_1}n_{r_2}
e^{iq(r_1-r_2)}\right>
=\frac 1{{\cal N}_{\ell}}\langle g(-q)g(q)\rangle=
\left<\left\{e^{iq(m_1+1)}+e^{iq(m_1+m_2+2)}+\dots\right\}\right.
\nonumber
\end{eqnarray}
\begin{eqnarray}
\left.\times\!\left\{e^{-iq(m_1+1)}+e^{-iq(m_1+m_2+2)}+\dots\right\}\right>
\end{eqnarray}
To evaluate Eq.(\ref{30}) we introduce the auxilary function:
\begin{eqnarray}
\label{31}
\Phi_k=\left<\left\{e^{iq(m_1+1)}+e^{iq(m_1+m_2+2)}+
\dots +e^{iq(m_1+m_2+\dots +m_k+k)}\right\}\right.
\nonumber
\end{eqnarray}
\begin{eqnarray}
\left.\times\!\left\{e^{-iq(m_1+1)}+e^{-iq(m_1+m_2+2)}+
\dots +e^{-iq(m_1+m_2+\dots +m_k+k)}\right\}\right>
\end{eqnarray}
Let us define $\,\xi=\langle e^{iq(m+1)}\rangle$.  Evidently,
$\,|\xi|\!<\!1\,$ for not a regular distribution of CFs in a chain.  Now
$\,\Phi_k$ can be rearranged as
\begin{eqnarray}
\label{32}
\Phi_k=\left(1+\xi^*+\dots +(\xi^*)^{k-1}\right)+
\left(\xi+1+\xi^*+\dots +(\xi^*)^{k-2}\right)+
\dots
\nonumber
\end{eqnarray}
\begin{eqnarray}
+\left(\xi^{k-1}+\xi^{k-2}+\dots+\xi+1\right)
=k+(k-1)(\xi+\xi^*)+\dots+(\xi^{k-1}+(\xi^*)^{k-1})
\nonumber
\end{eqnarray}
\begin{eqnarray}
=k\left(\frac{1-\xi^k}{1-\xi}-\frac 12\right)+
\frac{-\xi+k\xi^k-(k-1)\xi^{k+1}}{(1-\xi)^2}+c.c.
\end{eqnarray}
Going to the $\,k\!\rightarrow\!\infty\,$ limit according to the rule
$\,\lim_{{\cal N}_{\ell}\rightarrow\infty}k/{\cal N}_{\ell}=S\,$ we
obtain
\begin{eqnarray}
B(q)=S\left(\frac 1{1-\xi}+\frac 1{1-\xi^*}-1\right)
\nonumber
\end{eqnarray}
which is simultaneously Eq.(\ref{27}).

Value $B(q\!=\!0)$ is a matter of interest, because it enters the
thermally induced interaction terms of the high-temperature expansion of
the second order (see Appendix B). A small-$q$ expansion of $B(q)$
yields a direct reminiscence to the Lorenzian shape of the structure
factor:
\begin{eqnarray}
\label{34}
B(q)=S\left\{\frac {Sx^{(2)}}{\displaystyle 1+\frac {q^2}{12}
\left(3(x^{(2)})^2 -4x^{(3)}+\frac{x^{(4)}}{x^{(2)}}\right)}-1\right\}
\end{eqnarray}
where
$$
x^{(n)}=\sum_{m=0}(m+1)^nx_m
$$
Hence,
\begin{eqnarray}
\label{35}
B(q\rightarrow 0)\rightarrow S(Sx^{(2)}-1).
\end{eqnarray}
This expression can be simply evaluated at $\,S\!\ll\!1$.
Actually, according to Eqs.(\ref{18}),(\ref{20},(\ref{21})
\begin{eqnarray*}
Sx^{(2)}=S\frac{5!!\sqrt{2\pi m_0}m_0^3}{3!!\sqrt{2\pi m_0}m_0^2}=
5Sm_0=\frac 53
\end{eqnarray*}
In this case $B(q\rightarrow 0)\rightarrow 2S/3$. It tends to increase
with increasing $S$:
\begin{eqnarray}
\label{36}
B(q\rightarrow 0)\rightarrow S^2x^{(2)}\approx 5S^2(1-S)m_0
\end{eqnarray}
where $\,m_0$ is supposed to be large and temperature dependent (see
Eq.(\ref{20})).

The prefactor in the $q^2$-term of Eq.(\ref{34}) can be interpreted as
the square of the in-chain correlation length, $\lambda^2$, (positive or
negative!).  So, after doing a simple, but tedious algebraic exercise,
one obtains:
\begin{eqnarray}
\label{37}
|\lambda^2|=\left|\frac {25}4(1-S)^2-\frac {35}3(1-S)+\frac {21}4\right|m_0^2
\end{eqnarray}
The interesting feature of Eqs.(\ref{34}) and (\ref{37}) is that the
central peak at $q=0$ splits when $1-S>(14-{\sqrt 7})/15\approx\!0.757$.

Evidently, the knowledge of $B(q)$ allows to perform estimation of
the irreducible density-density correlation function:
\begin{eqnarray}
\label{38}
\langle\langle n_m n_{m+r}\rangle\rangle=\langle n_m n_{m+r}\rangle-
\langle n_m \rangle^2=\int_{-\pi}^{\pi}\frac{dq}{2\pi}B(q)e^{iqr}
\end{eqnarray}
Note that the important sum,
$\sigma(q)=S^{-1}e^{iq}\sum_{m=0}x_me^{iqm}$, entering $B(q)$, can be
asymptotically evaluated in the case of oxygen-rich chains
($m_0\!\gg\!1$).  It can be transformed in the Eq.(\ref{18}) manner:
\begin{eqnarray}
\label{39}
\sigma(q)=e^{iq}\frac{1+\sqrt{2\pi}e^{-\Phi}m_0^{3/2}(1-2iqm_0)^{-3/2}}
{1+\sqrt{2\pi}e^{-\Phi}m_0^{3/2}},
\end{eqnarray}
so, the general expression of $B(q)$ takes the form:
\begin{eqnarray}
\label{40}
B(q)=S\alpha\,\frac{2(1-u\cos\psi)+\alpha(1-u^2)}{2(1\!-\!\cos q)(1\!+
\!\alpha(1\!+\!u\cos\psi))+2u\alpha\sin\psi \sin q+\alpha^2(1+\!u^2-\!2u
\cos(q\!+\!\psi))}
\end{eqnarray}
where $\alpha=\sqrt{2\pi}m_0^{3/2}e^{-\Phi}$, $\,u=(1+4q^2m_0^2)^{-3/4}\,$
and $\psi=\frac 32\arctan 2qm_0$. One notes that by using Eq.(\ref{20})
all the parameters, entering Eq.(\ref{40}), can be expressed through
$m_0$. Namely, $\alpha=(1-S)/(Sm_0)$ and small. In the
$S\!\rightarrow\!0$-limit $\alpha$ is no more small, on the contrary,
$\alpha\!\gg\!1$ and $B(q)$ takes the form:
\begin{eqnarray}
\label{41}
B(q)=S\,\frac{1-u^2}{1+u^2-2u\cos(q+\psi)}
\end{eqnarray}
\section{Structure factor of interacting CF\lowercase{s}.}
In the preceding section we were interested in the behavior of the chain
fragment ensemble: {\it First,} in a hypothetical situation provided a
thermal equilibrium of oxygen in a basal plane is available at rather
low temperatures, {\it second,} in more realistic case when the
configurational entropy is not neglected.  The results obtained in the
latter case seem to be reasonable as regards the in-chain correlation
functions, although the interaction between the chains has been ignored.
However, the role of such an interaction is important in many physical
quantities which are sensitive to distinctions between Ortho-I and
Ortho-II phases, such as concentration of holes leaving CFs, $n_{\text h}$,
concentration of paramagnetically active CFs, $n_{\text {pm}}$, and
so on.  In this section we derive the formulas, describing that part of
diffuse scatterring caused directly by oxygen atoms and vacancies.  We
apply these formulas in numerical calculations, and visualize them in
Section IV.

Treating the problem in terms of the set $\{x_k\}$, we outline in
Appendix B the method which incorporates the interaction between nearest
chains into some scheme of statistical mechanics: This allows to
determine the correlated set $\{x_k^{\alpha}\}$, depending of the type
of orthorhombic phase.  The goal of this section is to develop a similar
approach for correlation functions, in particular, for the irreducible
part of the oxygen density-density correlation function.  The latter is
associated with the intensities of diffuse scatterring due to a partial
disorder in the CF ensemble:
\begin{eqnarray}
\label{42}
{\cal B}({\bf q})=\frac 1{\cal N}\sum_{{\bf r}_1,{\bf r}_2}G({\bf r}_1,
{\bf r}_2)
e^{i{\bf q}({\bf r}_1-{\bf r}_2)}
\end{eqnarray}
where
\begin{eqnarray}
\label{43}
G({\bf r}_1,{\bf r}_2)=
\langle\langle n_{{\bf r}_1}n_{{\bf r}_2}\rangle\rangle=
\langle n_{{\bf r}_1}n_{{\bf r}_2}\rangle
-\langle n_{{\bf r}_1}\rangle\langle n_{{\bf r}_2}\rangle
\end{eqnarray}
and ${\cal N}={\cal N}_{\ell}{\cal N}_{{\rm ch}}$.  It is evident, that
for ${\bf r}_1$ and ${\bf r}_2$, both belonging to the same chain, the
main order contribution into $G({\bf r}_1,{\bf r}_2)$ relates to the
in-chain correlation function discussed in subsection II.C.  Provided
${\bf r}_1$ and ${\bf r}_2$ are separated by $\,n\,$ chains, the main
order contribution appears to be
\begin{eqnarray}
\label{44}
G(\alpha,\mu_0;\alpha+n,\mu_n)=\left(-\frac {V_3}T\right)^n\sum_{m_0}\dots
\sum_{m_{n-1}}
\langle\langle n_{\mu_0}^{\alpha}n_{m_0}^{\alpha}\rangle\rangle
\,\langle\langle n_{m_0}^{\alpha+1}n_{m_1}^{\alpha+1}\rangle\rangle\cdots
\langle\langle n_{m_{n-1}}^{\alpha+n}n_{\mu_n}^{\alpha+n}\rangle\rangle
\end{eqnarray}
or, Eq.(\ref{44}) can be read as
\begin{eqnarray}
\label{45}
G_{\alpha,n}(q_y)=\frac 1{{\cal N}_{\ell}}\sum_{\mu_0,\mu_n}
\!G(\alpha,\mu_0;\alpha\!+\!n,\mu_n)e^{iq_y(\mu_0-\mu_n)}\!=\!
\left(-\frac{V_3}T\right)^n\!B_{\alpha}(q_y)B_{\alpha+1}(q_y)\dots
B_{\alpha+n}(q_y)
\end{eqnarray}
As regards the Eq.(\ref{45}) derivation we refer the reader to Appendix
C. We use in Eq.(\ref{45}) the $\,y$ component of the wave vector
following the convention about predominant orientation of chains along
$y$-axis.

Let us perform summation in ${\cal B}({\bf q})$: For any separation
$\,n\,$ between the chains the leading term of the order $(V_3/T)^n$
will be only kept.

The Ortho-I case is rather simple: All the elements $B_{\alpha}=B(q_y)$
in Eq.(\ref{45}) are $\alpha$-independent.  Summation over all the
elements $G_n(q_y)$ in Eq.(\ref{42}) yields:
\begin{eqnarray}
\label{46}
{\cal B}({\bf q})=B\left\{1+2\sum_{n=0}\cos nq_x\left(
-\frac{V_3}TB\right)^n\right\}=
B\left\{\frac 1{\displaystyle 1+\frac{V_3}Te^{iq_x}B}+
\frac 1{\displaystyle 1+\frac{V_3}Te^{-iq_x}B}-1\right\}
\end{eqnarray}
In the case of the Ortho-II phase the analogous expression is more
complex ($B_{\alpha}$ alters from chain to chain, taking the values,
say, $B_1(q_y)$ and $B_2(q_y)$ for oxygen-rich and oxygen-poor chains,
respectively):
\begin{eqnarray}
\label{47}
{\cal B}({\bf q})=\left\{\frac{B_1+B_2}2-
2\frac{V_3}TB_1B_2\cos q_x+
+\left(\frac{V_3}T\right)^2
(B_1+B_2)B_1B_2\cos 2q_x\right.
\nonumber
\end{eqnarray}
\begin{eqnarray}
\left.-2\left(\frac{V_3}T\right)^3
B_1^2B_2^2\cos 3q_x+\dots\right\}
\!=\!\left(\frac{\displaystyle \frac{B_1+B_2}2-\frac {V_3}Te^{iq_x}B_1B_2}
{\displaystyle 1-\frac{V_3^2}{T^2}e^{i2q_x}B_1B_2}+{\text c.c.}\right)
-\frac{B_1+B_2}2
\end{eqnarray}

Anticipating the results of numerical calculations let us find the
${\cal B}({\bf q})$ shape in the Ortho-I phase in the vicinity of the
Ortho-II Bragg peak ($\frac 12=\delta q_x,\delta q_y$): This point of
the reciprocal space is a suitable candidate to be used in determination
of correlation lengths along and across the chains.  One simply checks
that Eq.(\ref{46}) can be written as
\begin{eqnarray}
\label{48}
{\cal B}({\bf q})\!=\!B(\delta q_y)\frac{
\displaystyle 1-\left(\frac{V_3B(\delta q_y)}T\right)^2}
{\displaystyle 1-2\frac{V_3B(\delta q_y)}T\cos\delta q_x+
\left(\frac{V_3B(\delta q_y)}T\right)^2}
\end{eqnarray}
The correlation length across the chains is associated with the
denominator in Eq.(\ref{48}):
\begin{eqnarray}
\label{49}
\lambda_x^2=\frac{\frac{V_3B(0)}T}
{\left(1-\frac{V_3B(0)}T\right)^2}
\end{eqnarray}
so, $\lambda_x^2$ is positively defined.  The $\lambda_y^2$ expression
can be also easily obtained from Eq.(\ref{48}):
\begin{eqnarray}
\label{50}
\lambda_y^2/\lambda^2=\frac{2\frac{V_3B(0)}T}
{1-\left(\frac{V_3B(0)}T\right)^2}+1
\end{eqnarray}
For $\lambda^2$ see Eq.(\ref{37}).

If $V_3B(0)/T\!\ll\!1$, ie, provided the high-temperature expansion
holds, $\lambda_x$ is of the order of the lattice constant, whereas
$\lambda_y\sim |\lambda|\sim m_0$.
\section{Numerical calculations.}
Below some analytical results of the preceding section are visualized
after numerical calculations.  In order to make a selection of the
fitting parameters for the theory presented above we refer to the
plateau-like behavior of $n_{\text h}$ vs $x$ which should imitate the
60 K plateau of $T_c$ vs $x$ in YBCO.  In Fig.3 the $n_{\text h}(x)$
dependence is shown at $\tilde V_2=V_3=1$ and $T_q$, varying with
interval 0.08 from 0.46 to 0.86, which seems a reasonable choice for
interpretation of experimental facts.  The regions of stabilities of
Ortho-I and Ortho-II phases can be clearly distinguished in the set of
curves of Fig.3.  The curves, illustrating the monovalent copper amount
vs $x$ are shown in Fig.4. {\it \`A propos}, monovalent copper atoms
occupy 2-fold coordinated sites, which are out of CFs. This quantity,
$n_{\text{2-fold}}$, is certainly not in a contradiction with the
experimental data of Refs \cite{-21,-22}.  Closely related to
$n_{\text{2-fold}}$ are the concentrations of copper ions in 3- and
4-fold coordinated positions, ie, copper atoms at the ends of CFs and
within them, respectively (see Figs.5,6).  According to \cite{-29} the
information about distribution of copper atoms in differently
coordinated positions can be obtained in NQR experiments.  The quantity,
which also reflects the structural peculiarities of the Ortho-I and
Ortho-II phases, is the average length of chain fragments shown in
Fig.7.  The last physical quantity of the reference set available for
checking experimentally could be the concentration of paramagnetically
active chain fragments, $n_{\text {pm}}$.  The theoretical curves have
typical forms shown in Fig.8.  The experimental research program,
employing the polarized neutron technique is in the beginning (see
Ref.\cite{25}).  Also the concentration of paramagnetic CFs at low
temperatures could be measured due to the Shottky effect.  Below we give
shortly the explanations of how the afore-said concentrations can be
evaluated in the frame of the generalized lattice-gas model.
\subsection{Concentration of holes quitting the chains.}
The general expression of the concentration of holes transferred from
chains has a form:
\begin{equation}
\label{51}
n_{\text h}=\frac 12\sum_{n=1}\sum_{m=0}^n(m-1)<x_n>{\rm Pr}(n,m)
\end{equation}
Factor $\frac 12$ enters the equation because one oxygen deficient plane
supplies holes into two CuO$_2$ planes.  Function Pr($n,m$) is the
probability for a chain fragment of the length $n$ to have $m$ O$^{2-}$
ions within and can be obtained in the frame of some strongly correlated
model.  Here we accept a simplified consideration according to which the
most probable configuration has to be substituted Eq.(\ref{51}).  For
very long chains the most probable $\overline m$ at fixed $n$ is simply
$\xi n$, where $\xi=\lim_{n\rightarrow\infty}\xi_n$.  Although for
shorter chain fragments the relationship between $\overline m$ and $n$
is more delicate, it could be approximately expressed as
\begin{equation}
\label{52}
\overline {m(n)}\approx {\rm nint}(\xi n)
\end{equation}
where function {\it nint} denotes the nearest integer number.
So, in this case we rewrite Eq.(\ref{51}) as follows:
$$
n_{\text h}=\frac 12\sum_{n=1}({\rm nint}(\xi n)-1)\langle x_n\rangle
$$
\subsection{Concentration of paramagnetically active CF\lowercase{s}.}
Again, starting with some microscopical model for a finite CF
characterized by $n$ and $m$, one obtains, in principle, the average
value of $\vec S^2$.  To simplify the scheme (see also Ref.\cite{23}) we
first recall that there is no charge transfer from short chain clusters
with one or two oxygen atoms, both are spin-singlet fragments.  For a
cluster of three oxygen atoms, $n=3$, a spin-1/2 state forms,
contributing to the paramagnetic Curie constant.  On the other hand, for
$n=4$, one expects again a singlet state, because the most probable is
the $m=3$ state.  In the longer clusters, $n\ge 5$, singlet and doublet
states occur with roughly the same probability, at sufficiently high
temperatures.  So, the paramagnetic Curie constant is proportional to
the concentration of paramagnetically active CFs, $n_{\text {pm}}$,
which may be approximated by
\begin{equation}
\label{53}
n_{\text {pm}}=\langle x_3\rangle+\frac 12\sum_{n=5}\langle x_n\rangle
\end{equation}
\subsection{Cu$^+$ amount.}
The experiments \cite{-21,-22} were very important for understanding of
the charge transfer mechanism in YBCO.  Because only those copper site
which are surrounded by oxygen vacancies give a contribution into the
monovalent copper amount, their concentration coincides with
$\langle x_0\rangle$.
\subsection{Average lengths of CF\lowercase{s} and concentrations of Cu
atoms in 3- and 4-fold coordinations.}
Acceptable in applications to the Ortho-II phase are the formulas given
below and expressed by means of the $x_0$ values:
\begin{eqnarray}
\label{52.1}
\overline {\ell_c}=\frac{x^{(c)}}{S^{(c)}-x_0^{(c)}}
\end{eqnarray}
where ($c$) relates to either rich ($r$) or poor ($p$) chains;
their average characteristic takes a form:
\begin{eqnarray}
\label{52.2}
\overline {\ell}=\frac x{S-\langle x_0\rangle}
\end{eqnarray}
Noting that 2($N_1+N_2+\dots$) is the amount of copper in the 3-fold
coordination in a chain, one easily obtains $n_{\text{3-fold}}$
\begin{eqnarray}
\label{52.3}
n_{\text{3-fold}}=2-2x-2\langle x_0\rangle.
\end{eqnarray}
$n_{\text{4-fold}}$ can be simply derived due to a sum rule:
\begin{eqnarray}
\label{52.4}
n_{\text{4-fold}}=1-n_{\text{2-fold}}-n_{\text{3-fold}}=
2x-1+\langle x_0\rangle.
\end{eqnarray}
\subsection{Structure factor: Diffuse scatterring on the CF ensemble.}
Figs.9(a-f) illustrate a diffuse part of the structure factor calculated
numerically according to the scheme developed in this paper.  Remind,
that this part is due to structural imperfections in the CuO$_x$ plane.
The fitting parameters, $\tilde V_2$ and $V_3$, are common with the
values used for the plots in Figs.3-8.  Calculations were performed at
$T_q=0.62$: Figs.3-8 also include the curves relating to this
temperature.

The following features of the structure factor shape are necessary to be
mentioned:
\begin{itemize}
\item
Its value enhances on the Brillouin zone boundary, ($\pm 1/2,q_y$) (in
units of 2$\pi$), it reflects a tendency for a system to be arranged as
a double period structure on a short wavelength scale.  This fact would
become especially pronounced if the Ortho-I phase were forced to
substitute a true Ortho-II phase -- in this case a short-range oxygen
atom -- oxygen vacancy correlation across the chain should be
significant.
\item
The Ortho-I phase exhibits a clear down-fall behavior of the structure
factor in the ($\pm 1/2,0$) points of the reciprocal space. This reminds a
central peak splitting discussed in subsection II.C.
\item
The maximum diffuse scattering intensity is displayed in the vicinity
of the Ortho-I-- Ortho-II phase transition.
\end{itemize}
\section{Conclusive remarks and summary.}
This paper is devoted to the correlations in the oxygen deficient planes
which are the important structural elements of HT superconductors,
YBa$_2$Cu$_3$O$_{6+x}$ and related compounds.

There were many experiments performed to reveal the fascinating
properties of such compounds.  Some of them are quoted in this paper in
connection with superstructure formations, charge transfer mechanism,
magnetic behavior of CFs. There were also many attempts to describe
CuO$_x$ planes theoretically.  Many of them invoked into consideration
the oxygen subsystem only.  We already mentioned the conventional
lattice-gas approach \cite{5,6,15,16,18,19}; the approach based on the
screened Coulomb potential \cite{30} has been also applied to the oxygen
subsystem in order to explain some hypothetical superstructures.  The
model utilized in this paper looks as generalization of the lattice-gas
model (see also \cite{23}), but the role of copper in formation of
alternating Cu-O CFs and charge transfer mechanism is not ignored.

The statement, utilized in this paper, concerns the copper-oxygen chain
fragments: Many phenomena in YBCO are caused by them.  The function of
the CF ensemble to control the YBCO behavior is closely connected with
the charge transfer which could be conditionally subdivided into two
parts: the intrachain charge redistribution and that one from chains
onto CuO$_2$ planes.  The former is the oxydation of monovalent Cu ions,
forming the CFs, to the bivalent state According to this process almost
all oxygen atoms in not very short Cu-O CFs would be subjected to
transformation into the monovalent state where it not for the hole
transfer to outside: Around 30\% of oxygen atoms in chains persist their
monovalency, the rest oxygen constituents appear to be in the bivalent
state \cite{20}.  It is noteworthy, that the distribution of oxygen
valencies is neither frozen, nor ordered in any charge density
superstructure, but due to strong $p$-$d$ hybridization oxygen holes
form strongly correlated systems of finite lengths as discussed in
\cite{21}.

According to the detailed investigation by Jorgensen {\it et al}
{}~\cite{22} of the orthorhombically arranged CuO$_x$ planes the
concentration of oxygen atoms in "wrong" O(5) positions is negligibly
small up to rather high temperature $\sim$ 600$^{\circ}$C.  The
situation reminds the soft matter: Cu-O fragments behave as long
molecules strongly oriented and weakly correlated to each other in
chains.

Any oxygen migration stops approximately at and below room temperature.
So, this circumstance does not allow to appear those CF structures whose
existence could be a result of a simple energy competition.  Actually,
distribution described by Eq.(\ref{15}) has a maximum at $m_0$, but its
width is of the order 2$m_0$, hence, the role of the entropy becomes
decisive.

Below we resume the method and results presented in the paper.
\begin{itemize}
\item
The principal variables over which the statistical averaging can be done
are the probabilities, $\{x_m^{\alpha}\}$.
\item
The broad distribution of CFs around 3$m_0$ (see Eq.(\ref{23})) can be
utilized for calculation of such quantities as the hole transfer from
chains, concentration of paramagnetically active chain fragments, the
monovalent copper amount, etc. They evidently depend on the oxygen
content, $x$, and quenching tempearature, $T_q$.
\item
The in-chain structure factor can be used for interpretation of the
diffuse scatterring results in the Ortho-I phase.  Splitting of the 1d
central peak could be a counterpart of the analogous splitting of the
diffuse maximum on the Brillouin cell boundary, ($\pm 1/2,0$).
\item
Oxygen-oxygen correlations in the basal plane are illustrated in Figs 9.
The maximum of diffuse scatterring corresponds to the
Ortho-II-to-Ortho-I transition concentration.
\item
Measurements of the correlation length along the chains can be employed
for experimental determination of effective coupling constants.
\end{itemize}
The message of this work is clearly addressed to those experimentalists
whose field of interests lies in atomic and magnetic correlations
in YBCO and related compounds. Neutron and X-ray studies seem to be the
most appropriate technique for this. Nevertheless, the method developed
in the paper is possible for evaluation of thermodynamical quantities,
such as, isothermical susceptibility, Shottky anomaly, etc.
\vspace{0.5cm}

\subsection*{Acknowledgement.}
I thank the L\'eon Brillouin Laboratory at Saclay for its hospitality
during the initial stage of this work. I keep in my memory a deep interest
to the problem by J.Rossat-Mignod, deceased last August. I thank also
V.Plakhty, L.-P.Regnault and J.Schweizer for the  discussions which gave
rise to this work.

I want to acknowledge Prof.  J.Zittartz for his kind invitation to visit
the Institute for Theoretical Physics at the Cologne University where
the work has been completed.
\vfill
\newpage
\renewcommand{\theequation}{A.\arabic{equation}}
\setcounter{equation}{0}
\begin{appendix}
{\bf Appendix A.}
\vspace{0.5cm}

\noindent
Taking the formal derivatives of $f\{i,j\}$ (see Eq.(\ref{11})) over
$\,i\,$ and $\,j\,$ we get
\begin{eqnarray}
\label{A1}
\frac{\partial f\{i,j\}}{\partial j}=-\tau (1-x)\frac{\kappa-i}{(j-i)^2}
\left(\frac {j-i}j-\ln\frac ji\right)
\end{eqnarray}
\begin{eqnarray}
\label{A2}
\frac{\partial f\{i,j\}}{\partial i}=-\tau (1-x)\frac{j-\kappa}{(j-i)^2}
\left(\frac {j-i}i-\ln\frac ji\right)
\end{eqnarray}
Substituting $\,z=i/j<1$ we notice that the sign of derivative (\ref{A1})
coincides with the sign of expression $\,z-1-\ln z\,$ and positive. Let
$\,i\,$ be fixed, hence, $\,f\{i,j\}\,$ approaches the minimum as a
function of $\,j\,$ at the greater nearest integer to $\,\kappa$, say,
$\kappa_+$. The same substitution makes the sign of derivative (\ref{A2})
identical to the sign of expression $\,1-\ln z -1/z\,$ which is negative.
The minimum of $\,f\{i,j\}\,$ as function of $\,i\,$ is situated at the lower
nearest integer to $\,\kappa$, say, $\kappa_-$. Hence, $\,f\{i,j\}\,$ attains
the global minimum on the manifold of vertices, defined by Eq.(\ref{9})
with $0<i<j$, on unique vertex $\,[\kappa_-,\kappa_+]$.

In order to search, what kind of vertex, $\,[\kappa_-,\kappa_+]$ or
$[0,m]$, will be realized as a proper one, we start with the case
$\kappa_-\!>\!\mu=\exp( 1+u/\tau)$.  Functions
\begin{eqnarray*}
f_1=u-\tau\big((\kappa-\kappa_-)\ln \kappa_++(\kappa_+-\kappa)\ln \kappa_-
\big)\;\;\text {and}\;\;f_2=\frac {\kappa}m(u-\tau\ln m)
\end{eqnarray*}
differ from those of Eqs.(\ref{12}) and (\ref{13}) by a common
multiplayer.  Because in this case $\,m=\kappa_+$ is optimal for $f_2$
(see Eq.(\ref{13}),
\begin{eqnarray}
\label{A3}
f_1-f_2=u\left(1-\frac {\kappa}{\kappa_+}\right)-\tau\left(
(\kappa-\kappa_-)\ln \kappa_++(\kappa_+-\kappa)\ln \kappa_--
\frac {\kappa}{\kappa_+}\ln \kappa_+\right)
\nonumber
\end{eqnarray}
\begin{eqnarray}
=\tau\frac {\kappa_+-\kappa}{\kappa_+}(\ln \mu-1-\kappa_+\ln\kappa_-
-\kappa_+\ln\kappa_+)
\end{eqnarray}
appears to be negative and favors the $[\kappa_-,\kappa_+]$ vertex.
When deriving Eq.(\ref{A3}) we used the relation $\,\ln \mu=u/\tau+1$.
Contraversially, if $\kappa_+<\mu$ we substitute $f_2$ by
$\overline{f_2}$:
$$
f_2\rightarrow \overline{f_2}=\frac {\kappa}{\kappa_+}(u-\tau\ln \kappa_+)
\geq f_2
$$
Then
\begin{eqnarray}
\label{A4}
f_1-f_2\geq f_1-\overline{f_2}
=\tau\frac {\kappa_+-\kappa}{\kappa_+}(\ln \mu-1-\kappa_+\ln\kappa_-
-\kappa_+\ln\kappa_+)>0
\end{eqnarray}
To complete the proof we consider the last possibility,
$\kappa_-\!<\!\mu\!<\!\kappa_+$.  Again $\kappa_+$ substitutes for $\mu$
and the sign of ($f_1-f_2$) coincides with the r.h.s. of Eq.(\ref{A3})
which turns to zero at
$$
\mu\approx \kappa_-+\frac 12-\frac 5{24\kappa_-},\;\;\kappa_-\gg 1
$$
\end{appendix}
\vspace{1cm}

\renewcommand{\theequation}{B.\arabic{equation}}
\setcounter{equation}{0}
\begin{appendix}
{\bf Appendix B.}
\vspace{0.5cm}

\noindent
Deriving the equations of statistical mechanics for oxygen in a deficient
plane we suppose:
\begin{itemize}
\item
A negligible probability for oxygen to occupy a position across the
chains;
\item
The largest energy contribution comes from the interchain degrees of
freedom;
\item
Apart from the entropy contribution to the total energy of the CF
ensemble we take into account the oxygen-oxygen repulsion on the nearest
chains, like in the lattice-gas model (cf, \cite{5,6,19}).
\item
The constraint used is the total oxygen content, $x$ (see Eq.(\ref{7})).
\end{itemize}
As regards the second supposion it denotes a preferable role of strong
fermionic correlations within $\dots$-Cu-O-$\dots$ chain fragments as
compared with interchain interactions which are effectively screened in
the metallic state of YBCO.  The oxygen ordering in deficient planes
reminds the soft matter case, e.g., the nematic liquid crystal.  Whereas
the CFs weakly correlate to each other they are strongly coupled from
inside and preferably oriented in chains.

Let $\{N^{\alpha}_0,N^{\alpha}_1,N^{\alpha}_2,\dots,
N^{\alpha}_k,\dots\}$ be the total numbers of CFs of various lengths
(for their definitions see the beginning of Section II) in the
$\alpha$-th chain.  A complete number of different configurations of all
those CFs has a form usual for a mixture of ideal gases:
\begin{eqnarray}
\label{B1}
\frac{\left(N^{\alpha}_0+N^{\alpha}_1+N^{\alpha}_2+\dots+
N^{\alpha}_k+\dots\right)!}
{N^{\alpha}_0!N^{\alpha}_1!N^{\alpha}_2!\dots
N^{\alpha}_k!\dots}
\end{eqnarray}
Before giving a definition of the partition function of the CF ensemble
in the oxygen deficient plane let us omit for a while the $V_3$-term and
consider the auxilary partition function, ${\cal Z}^{\alpha}_0$, of the
$\alpha$-th chain with two constraints imposed: (\ref{5}) and (\ref{6}).
Playing the role of the Lagrange multiplayers are the chemical
potentials, $\mu^{\alpha}$ and $\mu^{\alpha}_0$, conjugated to the above
mentioned constraints.  With such a prerequisite ${\cal Z}^{\alpha}_0$
takes a form:
\begin{eqnarray}
\label{B2}
{\cal Z}^{\alpha}_0=\sum_{\{x_k^{\alpha}\}}\exp {\cal N}_{\ell}\left\{
S_{\alpha}\ln S_{\alpha}-\sum_{m=0}x_m^{\alpha}\ln x_m^{\alpha}
-\frac 1T\sum_{m=1}m\phi(m)x_m^{\alpha}\right.
\nonumber
\end{eqnarray}
\begin{eqnarray}
\left.+\frac{\mu^{\alpha}}T\sum_{m=0}(m+1)x_m^{\alpha}-\frac{\mu_0^{\alpha}}T
S^{\alpha}\right\}=\sum_{\{x_k^{\alpha}\}}\exp A^{\alpha}(\{x_k^{\alpha}\})
\end{eqnarray}
where summation runs over the possible distribution of CF lengths,
$A^{\alpha}(\{x_k^{\alpha}\})$ is the corresponding "action".

The equilibrium distribution of "probabilities" $\{x_k^{\alpha}\}$ achieves
at the maximal "action" and can be expressed as follows:
\begin{eqnarray}
\label{B3}
x_k^{\alpha}=\frac {e^{-k(\phi(k)-\mu^{\alpha})/T}}{\displaystyle
\sum_{m=0}(m+1)e^{-m(\phi(m)-\mu^{\alpha})/T}}
\end{eqnarray}
\begin{eqnarray}
\label{B4}
S^{\alpha}=\frac {\displaystyle\sum_{m=0}e^{-m(\phi(m)-\mu^{\alpha})/T}}
{\displaystyle\sum_{m=0}(m+1)e^{-m(\phi(m)-\mu^{\alpha})/T}}
\end{eqnarray}
In principle, $\mu_0^{\alpha}$ can be also expressed through $\mu^{\alpha}$:
\begin{eqnarray}
\label{B5}
e^{\mu^{\alpha}_0/T}=e^{\mu^{\alpha}/T}\sum_{m=0}
e^{-m(\phi(m)-\mu^{\alpha})/T}
\end{eqnarray}
So, any ${\cal Z}^{\alpha}_0$ can be considered as a function of
$\mu_{\alpha}$.  Substituting $\phi(k)$ in Eqs.(\ref{B3})-(\ref{B4}) by
its approximation of Eq.(\ref{3}) one obtains functions
(\ref{15})-(\ref{17}).

Let us include the $V_3$-term into consideration:
\begin{eqnarray}
\label{B6}
{\cal Z}=\prod_{\{x_k^{\alpha}\}}e^{A^{\alpha}(\{x_k^{\alpha}\})}
\zeta^{\alpha,\alpha+1};\;\;\;\;\zeta^{\alpha,\alpha+1}=
\exp\left(-\frac{V_3}T\sum_{m+1}^{{\cal N}_{\ell}}n_i^{\alpha}n_i^{\alpha+1}
\right)
\end{eqnarray}
Then it is suitable to expand $\zeta^{\alpha,\alpha+1}$ over the
cumulants.  Such a scheme is well-known and we only outline it below
($\eta= \prod_{\alpha}\zeta^{\alpha,\alpha+1}$):
\begin{eqnarray*}
{\cal Z}=\sum_{\{x_k^{\alpha}\}}\dots\eta=
\sum_{\{x_k^{\alpha}\}}\dots\left(1+\eta+\frac 12\eta^2+\dots
\right)={\cal Z}_0\left(1+\langle\eta\rangle_0+\frac 12
\langle\eta^2\rangle_0+\dots\right)
\end{eqnarray*}
where $<\dots>_0$ an average over the ensemble of non-interacting chains,
${\cal Z}_0=\prod_{\alpha}{\cal Z}_0^{\alpha}$. For example,
\begin{eqnarray*}
\langle\eta\rangle_0=-\frac{{\cal N}_{\ell}V_3}T\sum_{\alpha}
\overline {n^{\alpha}n^{\alpha+1}}.
\end{eqnarray*}
The expectation values, entering the $\langle\eta\rangle_0$-term of the
high-temperature expansion, can be expressed through the $S$-values,
supposed to be known:
\begin{eqnarray*}
\sum_{\alpha}\overline {n^{\alpha}n^{\alpha+1}}=
\sum_{\alpha}\overline {n^{\alpha}}\;\overline {n^{\alpha+1}}=\sum_{\alpha}
(1-S^{\alpha})(1-S^{\alpha+1})=(2x-1){\cal N}_{{\rm ch}}+\sum_{\alpha}
S^{\alpha}S^{\alpha+1}
\end{eqnarray*}
$\langle\eta^2\rangle_0$ consists of the expectation values of the forth
order, $\overline {n_i^{\alpha}n_i^{\alpha+1}n_j^{\beta}n_j^{\beta+1}}$,
which completely decouple if $\,\beta>\alpha+1\,$ or $\,\beta<\alpha-1$.
Also,
\begin{eqnarray}
\label{B7}
\overline {n_i^{\alpha-1}n_i^{\alpha}n_j^{\alpha}n_j^{\alpha+1}}
\rightarrow \overline {n^{\alpha-1}}\;\overline {n^{\alpha+1}}
\left(\overline {n^{\alpha}}^2+
\left<\!\left<n_i^{\alpha}n_j^{\alpha}\right>\!\right>_0\right)
\end{eqnarray}
where
\begin{eqnarray}
\label{B8}
\langle\langle n_i^{\alpha}n_j^{\alpha}\rangle\rangle_0=\overline
{n_i^{\alpha}n_j^{\alpha}}-\overline {n^{\alpha}}^2
\end{eqnarray}
According to a definition $\langle\langle n_i^{\alpha}n_j^{\beta}
\rangle\rangle_0\equiv 0$ at $\alpha\neq\beta$. Note, that Eq.(\ref{B8})
is invariant under the oxygen atom $\Leftrightarrow$ oxygen vacancy
transformation ($n_m^{\alpha}\leftrightarrow 1-n_m^{\alpha}$).

There is another non-trivial contribution of the $\eta^2$-term:
\begin{eqnarray}
\label{B9}
\overline {n_i^{\alpha}n_i^{\alpha+1}n_j^{\alpha}n_j^{\alpha+1}}
\rightarrow
\left(\overline {n^{\alpha}}^2+
\langle\langle n_i^{\alpha}n_j^{\alpha}\rangle\rangle_0\right)
\left(\overline {n^{\alpha+1}}^2+
\langle\langle n_i^{\alpha+1}n_j^{\alpha+1}\rangle\rangle_0\right)
\end{eqnarray}
All the decoupled terms, coming from all the orders of the
high-temperature expansion, can be recollected into
\begin{eqnarray}
\label{B10}
\exp\left(-\frac{{\cal N}_{\ell}V_3}T\sum_{\alpha}S^{\alpha}S^{\alpha+1}
\right).
\end{eqnarray}

The terms $\langle\langle n_i^{\alpha}n_j^{\alpha}\rangle\rangle_0
\overline {n^{\alpha-1}}\;\overline {n^{\alpha+1}}$ (see Eq.(\ref{B7}))
give rise a thermally induced interaction of the next-to-nearest chains.
The recollected contribution of Eq.(\ref{B10}) is of the main order in
the $V_3/T$ expansion: It formally corresponds to the mean-field
approach.  The $V_3$-exponential term (Eq.(\ref{B10})) simply
contributes ${\cal Z}_0$ and does not formally influence on
Eqs.(\ref{B3})-(\ref{B4}), but Eq.(\ref{B5}) changes its form
\begin{eqnarray}
\label{B11}
e^{\mu_0/T}=e^{\mu^{\alpha}/T}\sum_{m=0}
\exp -\frac{m(\phi(m)-\mu^{\alpha})}T\!\times\!
\exp -\frac{V_3(S^{\alpha-1}+S^{\alpha-1})}T
\end{eqnarray}

It is noteworthy, that Eq.(\ref{B11}) is a system of non-linear
equations which couple the densities of oxygen vacancies in nearest
chains.  Actually, $\,S^{\alpha}=S^{\alpha}(\mu^{\alpha})$ (see
Eq.(\ref{B4})), hence, the r.h.s. of Eq.(\ref{B11}) is some function of
$S^{\alpha-1}, S^{\alpha}, S^{\alpha+1}$, whereas the l.h.s. of
Eqs.(\ref{B11}) does not depend of index $\alpha$ and contains the
function of the {\it global} chemical potential, $\mu_0$, conjugated to
the {\it global} density of oxygen atoms (vacancies).
\end{appendix}
\vspace{1cm}

\renewcommand{\theequation}{C.\arabic{equation}}
\setcounter{equation}{0}
\begin{appendix}
{\bf Appendix C.}
\vspace{0.5cm}

\noindent
The correlation function entering the r.h.s. of Eq.(\ref{43}) can be
expressed through the partition function of non-interacting chains,
${\cal Z}_0$, as follows:
\begin{eqnarray}
\label{C1}
\langle n_{{\bf r}_1}n_{{\bf r}_2}\rangle=
\frac{\displaystyle\sum_{\{x_k\}}{\cal Z}_0n_{{\bf r}_1}n_{{\bf r}_2}
\exp\left(-\frac {V_3}T\sum_{\alpha,i}n_i^{\alpha}n_i^{\alpha+1}\right)}
{\displaystyle\sum_{\{x_k\}}{\cal Z}_0
\exp\left(-\frac {V_3}T\sum_{\alpha,i}n_i^{\alpha}n_i^{\alpha+1}\right)}=
\nonumber
\end{eqnarray}
\begin{eqnarray}
=\frac{\displaystyle\langle n_{{\bf r}_1}n_{{\bf r}_2}\rangle_0
-\frac {V_3}T\sum_{\alpha,i}\langle n_{{\bf r}_1}n_{{\bf r}_2}
n_i^{\alpha}n_i^{\alpha+1}\rangle_0
+\frac {V_3^2}{2T^2}\sum_{\alpha,i}\sum_{\beta,j}
\langle n_{{\bf r}_1}n_{{\bf r}_2}
n_i^{\alpha}n_i^{\alpha+1}n_j^{\beta}n_j^{\beta+1}\rangle_0+\dots}
{\displaystyle 1-\frac {V_3}T\sum_{\alpha,i}
\langle n_i^{\alpha}n_i^{\alpha+1}\rangle_0
+\frac {V_3^2}{2T^2}\sum_{\alpha,i}\sum_{\beta,j}
\langle n_i^{\alpha}n_i^{\alpha+1}n_j^{\beta}n_j^{\beta+1}\rangle_0+\dots}
\end{eqnarray}
In order to deal with irreducible averages one needs the
$\langle n_{{\bf r}} \rangle$ definition:
\begin{eqnarray}
\label{C2}
\langle n_{{\bf r}}\rangle=
\frac{\displaystyle\sum_{\{x_k\}}{\cal Z}_0n_{{\bf r}}
\exp\left(-\frac {V_3}T\sum_{\alpha,i}n_i^{\alpha}n_i^{\alpha+1}\right)}
{\displaystyle\sum_{\{x_k\}}{\cal Z}_0
\exp\left(-\frac {V_3}T\sum_{\alpha,i}n_i^{\alpha}n_i^{\alpha+1}\right)}=
\nonumber
\end{eqnarray}
\begin{eqnarray}
=\frac{\displaystyle\langle n_{{\bf r}}\rangle_0
-\frac {V_3}T\sum_{\alpha,i}\langle n_{{\bf r}}
n_i^{\alpha}n_i^{\alpha+1}\rangle_0
+\frac {V_3^2}{2T^2}\sum_{\alpha,i}\sum_{\beta,j}
\langle n_{{\bf r}}
n_i^{\alpha}n_i^{\alpha+1}n_j^{\beta}n_j^{\beta+1}\rangle_0+\dots}
{\displaystyle 1-\frac {V_3}T\sum_{\alpha,i}\langle n_i^{\alpha}
n_i^{\alpha+1}\rangle_0
+\frac {V_3^2}{2T^2}\sum_{\alpha,i}\sum_{\beta,j}
\langle n_i^{\alpha}n_i^{\alpha+1}n_j^{\beta}n_j^{\beta+1}\rangle_0+\dots}
\end{eqnarray}

It has been known since the beginning of sixties \cite{31} that in
Ising-like systems a correlation function can be expressed as a power
series over irreducible averages, the so-called cumulants.  A true
expansion must only contain connected diagrams.  In our particular
problem the interaction involves atoms of nearest chains, hence, a
series expansion of the irreducible average, $\langle\langle
n_i^{\alpha}n_j^{\alpha+n} \rangle\rangle$, starts with a term
proportional to $(V_3/T)^n$:
\begin{eqnarray}
\label{C3}
\langle\langle n_i^{\alpha}n_j^{\alpha+n}\rangle\rangle=
\left(\frac{V_3}T\right)^n
\sum_{m_1}\langle\langle n_i^{\alpha}n_{m_1}^{\alpha}\rangle\rangle_0
\sum_{m_2}\langle\langle n_{m_1}^{\alpha+1}n_{m_2}^{\alpha+1}\rangle\rangle_0
\cdots
\sum_{m_n}\langle\langle n_{m_n}^{\alpha+n}n_j^{\alpha+n}\rangle\rangle_0
\end{eqnarray}
\end{appendix}

\vspace{1.5cm}

{\Large Figure Captions.}
\vspace{.2cm}

\par\noindent
{\bf Figure 1.} The structural phase diagram is shown qualitatively.
The bottom is confined by
room temperature, $T_R$: Below $T_R$ oxygen is practically frozen.
Probably the Tetra-to-Ortho-II phase transition is of the 1st order and
the $T-x$ phase diagram contains the phase separation area.
Hypothetically, the superstructures of longer periodicities, like Ortho-III,
could appear between Ortho-II and Ortho-II areas.
\vspace{.2cm}

\par\noindent
{\bf Figure 2.} The lattice-gas model definition, involving the oxygen-oxygen
interactions only (circles and crosses are for oxygen and copper atoms,
respectively).
\vspace{.2cm}

\par\noindent
{\bf Figure 3.} $n_{\text h}$ vs $x$: Six curves are shown for the
equidistant
set of quenching temperatures, from 0.46 to 0.86; $\tilde V_2=V_3=1$;
The higher temperature, the smaller $n_{\text h}$.
\vspace{.2cm}

\par\noindent
{\bf Figure 4.} $n_{\text {Cu}^+}$ ($n_{\text{2-fold}}$) vs $x$:
Six curves are shown for the equidistant
set of quenching temperatures, from 0.46 to 0.86; $\tilde V_2=V_3=1$;
The higher temperature, the smaller $n_{\text{2-fold}}$.
\vspace{.2cm}

\par\noindent
{\bf Figure 5.} $n_{\text{3-fold}}$ (copper in 3-fold coordination) vs $x$:
Six curves are shown for the equidistant
set of quenching temperatures, from 0.46 to 0.86; $\tilde V_2=V_3=1$;
The higher temperature, the larger $n_{\text{3-fold}}$.
\vspace{.2cm}

\par\noindent
{\bf Figure 6.} $n_{\text{4-fold}}$ (copper in 4-fold coordination) vs $x$:
Six curves are shown for the equidistant
set of quenching temperatures, from 0.46 to 0.86; $\tilde V_2=V_3=1$;
The higher temperature, the smaller $n_{\text{4-fold}}$.
\vspace{.2cm}

\par\noindent
{\bf Figure 7.} Average length of CFs: $\ell$ (solid curves),
$\ell_r$ (dashed curves), $\ell_p$ (dotted curves); $\tilde V_2=V_3=1$;
$T_q=$ 0.46, 0.62, 0.78.
The higher temperature, the smaller $\ell$'s.
\vspace{.2cm}

\par\noindent
{\bf Figure 8.} $n_{\text {pm}}$ vs $x$: Six curves are shown for the
equidistant
set of quenching temperatures, from 0.46 to 0.86; $\tilde V_2=V_3=1$;
The higher temperature, the larger $n_{\text {pm}}$.
\vspace{.2cm}

\par\noindent
{\bf Figure 9.} The structure factor, ${\cal B}({\bf q})$, as defined in the
text. $\tilde V_2=V_3=1$; $T_q=0.62$; Oxygen content changes equidistantly
from $x=$ 0.45 (a) to 0.95 (f). Chain orientation is supposed along the
$y$-axis.

\begin{references}
\bibitem{13} Oxygen content $x=0.35$ reported in Refs.11-12 is contraversial
because of $T_N$ determined in the same experiments: $x=0.25$ fits better
the $T_N$ data (pointed by P.Burlet).
\end{references}
\end{document}